\documentclass[11pt,showpacs,preprintnumbers,amsmath,amssymb,prd,nofootinbib,superscriptaddress]{revtex4-2}

\usepackage{dcolumn}
\usepackage{bm}
\usepackage{ifpdf}
\usepackage{hyperref}
\usepackage{bm}
\usepackage{xcolor,color,graphicx,graphics,physics}
\usepackage[spanish,english]{babel}
\usepackage[latin1]{inputenc}
\usepackage[OT1]{fontenc}
\usepackage{latexsym,amssymb,amsmath,amsfonts, slashed}
\usepackage{makeidx}
\usepackage{epsfig,subfigure}
\usepackage{natbib}
\usepackage{epstopdf}
\usepackage{mathrsfs}
\usepackage{hyperref}
\hypersetup{colorlinks=true, linkcolor=blue, citecolor=blue, urlcolor=blue}
\usepackage{enumerate}

\usepackage{fixmath}

\everymath{\displaystyle}
\usepackage{graphicx}

\usepackage[T1]{fontenc}
\usepackage{amsmath}
\usepackage{amssymb}
\usepackage{graphicx}
\usepackage{xcolor}

\newcommand{\bea}{\begin{eqnarray}}
\newcommand{\eea}{\end{eqnarray}}

\newcommand{\orcid}[1]{\href{https://orcid.org/#1}{\includegraphics[width=10pt]{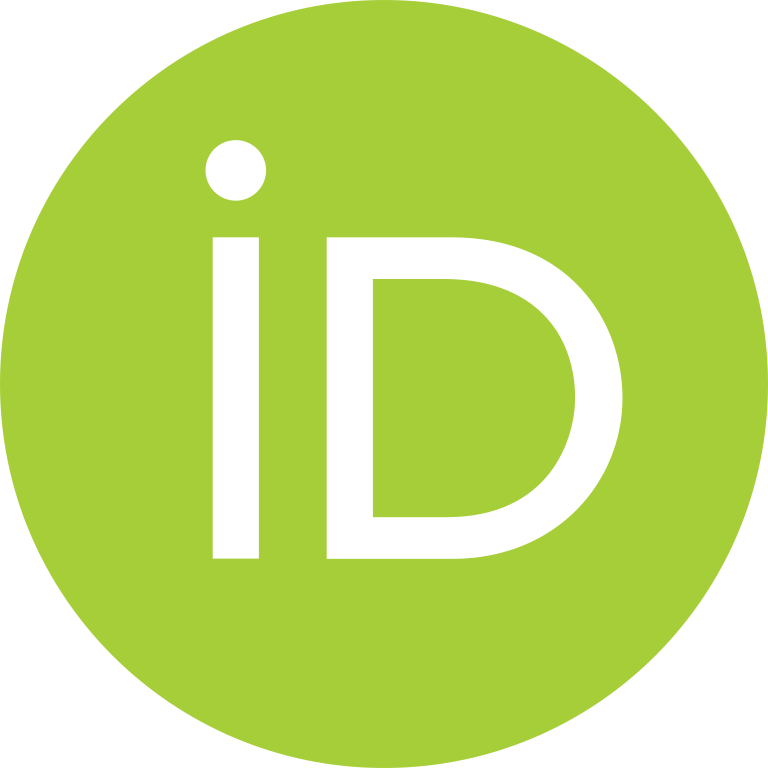}}}

\begin{document}

\title{Aether field coupled to the electromagnetic field in the TFD formalism}

\author{R. Corr\^ea}
\email{robson@fisica.ufmt.br}
\affiliation{Instituto de F\'{\i}sica, Universidade Federal de Mato Grosso,\\
78060-900, Cuiab\'{a}, Mato Grosso, Brazil}

\author{L. H. A. R. Ferreira \orcid{0000-0002-4384-2545}}
\email{luiz.ferreira@fisica.ufmt.br }
\affiliation{Instituto de F\'{\i}sica, Universidade Federal de Mato Grosso,\\
78060-900, Cuiab\'{a}, Mato Grosso, Brazil}

\author{A. F. Santos \orcid{0000-0002-2505-5273}}
\email{alesandroferreira@fisica.ufmt.br}
\affiliation{Instituto de F\'{\i}sica, Universidade Federal de Mato Grosso,\\
78060-900, Cuiab\'{a}, Mato Grosso, Brazil}

\author{Faqir C. Khanna \orcid{0000-0003-3917-7578} \footnote{Professor Emeritus - Physics Department, Theoretical Physics Institute, University of Alberta\\
Edmonton, Alberta, Canada}}
\email{fkhanna@ualberta.ca; khannaf@uvic.ca}
\affiliation{Department of Physics and Astronomy, University of Victoria,\\
3800 Finnerty Road, Victoria BC V8P 5C2, Canada}

\begin{abstract}

In this paper, the aether field, which leads to the violation of Lorentz symmetries, coupled with the electromagnetic field is considered. In order to study thermal and size effects in this theory, the Thermo Field Dynamics (TFD) formalism is used. TFD is a real-time quantum field theory that has an interesting topological structure. Here three different topologies are taken, then three different phenomena are calculated. These effects are investigated considering that the aether field can point in different directions. The results obtained are compared with the usual results of the Lorentz invariant electromagnetic field. 

\end{abstract}

\maketitle

\section{Introduction}

In recent decades, the violation of Lorentz symmetries has been  studied widely. The main idea that led to a quantum field theory with violation of Lorentz symmetries came from a context with extra dimensions in string theory \cite{KS,Pot}.  These studies lead to a question: is Lorentz invariance a fundamental or emergent phenomenon? Investigations that seek a unified theory point out that the Lorentz symmetry is an emergent symmetry. It is violated at very high energies and small traces should appear at low energies.  As a consequence, Planck scale symmetries may be different from the usual ones. Therefore,  in an emergent scenario, it is not obvious that Lorentz symmetry is implemented at the fundamental level, as is commonly seen. To investigate this symmetry break, an extension of known field theory models has been proposed \cite{Kostelecky3,Kostelecky4,Kostelecky5}. Among the various proposals that study the Lorentz violation, an interesting one is the aether field \cite{Carroll}. The first proposal consists of investigating hidden extra dimensions considering Lorentz violating tensor fields with expectation values along the extra directions. Interactions with other fields, such as scalar, gauge, and fermions fields, lead to modified dispersion relations. Several studies with the aether field, both in four dimensions and in extra dimensions, have been developed \cite{Ob,Ob1,Chat,Baeta,Mariz,Our,Capa,Carroll2,Eduardo}. In this paper, the objective is to study the aether field in four dimensions within the Thermo Field Dynamics (TFD) formalism.

The TFD formalism has a topological structure that allows the investigation of different phenomena on an equal footing  \cite{tfd1,khannatfd,Umezawa:1982nv,Umezawa:1993yq,lietfd}. Although TFD is known as a thermal quantum field theory, the temperature effect is not the only ingredient provided by this formalism. In addition to temperature effects,  size effects are also investigated in this topological theory. Then it is possible to calculate thermal effects as given in the Stefan-Boltzmann law and size effects as exhibited by the Casimir effect. However, to develop such an investigation, two fundamental elements are necessary: the Hilbert space is duplicated and the Bogoliubov transformation is used. The doubled Hilbert space or thermal Hilbert space allows the construction of a thermal ground state that leads to an important result, namely,  it is shown that the statistical average of an arbitrary operator is equal to its vacuum expectation value. 

The topological structure of the TFD is defined as $\Gamma_D^d=(\mathbb{S}^1)^d\times \mathbb{R}^{D-d}$ where $D$ are the space-time dimensions and $d$ is the number of compactified dimensions, with $1\leq d \leq D$. In this formalism, any set of dimensions of the manifold $\mathbb{R}^{D}$ can be compactified into a circle $\mathbb{S}^1$ which is specified by the parameter $\alpha_n$ known as the compactification parameter. Here, the TFD formalism is used for three different topologies, which implies that three different phenomena are investigated. The first topology consists of a temporal compactification, i.e., the time coordinate will be compactified into a circumference such that its length is related to the temperature. In this context, the Stefan-Boltzmann law for the aether field coupled to the electromagnetic field is calculated. In the second topology, the $z$ coordinate is compactified into a circle of length $L$. Here the size effects emerge, then the Casimir effect at zero temperature is studied. And in the last application, the considered topology allows both effects, i.e., the time and $z$ coordinates are compactified. Then, the Casimir effect at finite temperature is discussed. In all topologies, corrections due to the aether field to the usual electromagnetic results are analyzed.

This paper is organized as follows. In section II, the TFD formalism is briefly presented. In section III, the theory that describes the aether field coupled to the electromagnetic field is introduced. Using tools from the TFD formalism, the energy-momentum tensor associated with this model is calculated. In section IV, thermal and size effects applications are investigated. In addition, a vector constant that acts as a background field that is of time-like or space-like. In section V, some concluding remarks are discussed.

\section{Thermofield Dynamics Formalism} 

In this section, a brief introduction to the TFD formalism is presented. The TFD formalism is a quantum thermal field theory that allows us to write the statistical mean of an observable as its vacuum expected value, considering a thermal ground state. Its structure preserves the temporal information of the system so that it is possible to analyze the effects of temperature over time. For this reason, it is known as real-time formalism. In order to construct a thermal Hilbert space, it is necessary to introduce a dual or tilde Hilbert space, which implies a doubling of the system's degrees of freedom. Therefore, the thermal Hilbert space ${\cal H}_T $ is defined by the direct product of the original Hilbert space $ {\cal H} $ and the dual or tilde space $ \tilde{{\cal H}} $, i.e. $ {\cal H}_T={\cal H}\otimes\tilde{{\cal H}} $. To obtain the tilde variables, the tilde conjugation rules are used. These rules map the tilde and non-tilde variables and are written as
\begin{align}
	({\cal A}_i{\cal A}_j)^{\sim}&=\tilde{{\cal A}}_i\tilde{{\cal A}}j\nonumber\\
	(c{\cal A}_i+{\cal A}_j)^{\sim}&=c^*\tilde{{\cal A}}_i+\tilde{{\cal A}}_j\nonumber\\
	({\cal A}_i^\dagger)^\sim&=\tilde{{\cal A}}_i^\dagger\nonumber\\
	(\tilde{{\cal A}}_i)^\sim&=-\xi{\cal A}_i\label{1},
\end{align}
where $ {\cal A} $ is an arbitrary operator and $\xi=-1$ ($+1$ ) for bosons (fermions).

Another essential tool within this elegant formalism is the Bogoliubov transformation. This is characterized by the rotation of the tilde and non-tilde variables, which implies the introduction of a new parameter in these variables. This parameter is called the compactification parameter and is defined as $ \alpha=(\alpha_0,\alpha_1,..,\alpha_N) $. Considering a 4-dimensional space-time, if $\alpha=1/\beta$ with $\beta=1/k_B T$, for example, we have a description of the effects of temperature and the compactified dimension is the temporal one, from which the topology  $ \Gamma^1_4=\mathbb{S}^1\times\mathbb{R}^3 $ is obtained. Here, $T$ is the temperature and $k_B$ the Boltzmann constant.

As an example, two arbitrary operators $ {\cal A} $ and $ \tilde{{\cal A}} $ are considered. Then the Bogoliubov transformation leads to
\begin{align}
	\left(\begin{array}{c}
		{\cal A}(k,\alpha)\\
		\xi\tilde{{\cal A}}^\dagger(k,\alpha)
	\end{array}\right)={\cal B}(\alpha)\left(\begin{array}{c}
	{\cal A}(k)\\
	\xi\tilde{{\cal A}}^\dagger(k)
\end{array}\right),
\end{align}
where
\begin{align}
	{\cal B}(\alpha)=\left(\begin{array}{cc}
		u(\alpha)&-v(\alpha)\\
		\xi v(\alpha)&u(\alpha)
	\end{array}\right),
\end{align}
with $ u(\alpha) $ and $ v(\alpha) $ representing the Bose-Einstein distribution such that $ v^2(\alpha) =\dfrac{1}{e^{\alpha\omega}-1} $ and $ u^2(\alpha)=1+v^2(\alpha) $. Here, the propagator of the electromagnetic field is considered and the changes caused by the presence of the parameter $\alpha$ are analyzed. Such a propagator can be written as
\begin{eqnarray}
	iD_{\mu\nu}^{(ab)}(x-x',\alpha)&=&\langle 0,\tilde{0}|\tau[A^a_\mu(x;\alpha)A^b_\nu(x';\alpha)]|0,\tilde{0}\rangle\nonumber\\
	&=&\eta_{\mu\nu}G_0^{(ab)}(x-x';\alpha),\label{Prop}
\end{eqnarray}
where $ a,b=1,2 $, $\tau$ is the time ordering operator, $\eta_{\mu\nu}$ is the Minkowski metric and the field is defined in such a way that $A^a_\mu(x;\alpha)={\cal B}(\alpha)A^a_\mu(x){\cal B}^{-1}(\alpha)$. Note that, $G_0^{(ab)}(x-x';\alpha)$ is the scalar field propagator that depends on $\alpha$ which is written as
\begin{align}
	G_0^{(ab)}(x-x';\alpha)=\int\frac{d^4k}{(2\pi)^4}e^{-ik\cdot(x-x')}G^{ab}_0(k;\alpha),
\end{align}
with $G_0^{(ab)}(k;\alpha)={\cal B}(\alpha)G_0^{(ab)}(k){\cal B}^{-1}(\alpha)$. 

\par As physical quantities, that is, those that can be measured, are described by non-tilde variables, only these are considered. Then the Green function is given as
\begin{align}
	G_0^{(11)}(k;\alpha)=G_0(k)+\xi v^2(k;\alpha)[G^*_0(k)-G_0(k)],
\end{align}
where $ v^2(k;\alpha) $ is the generalized Bogoliubov transformation defined as
\begin{align}
	v^2(k;\alpha)=\sum_{s=1}^{d}\sum_{\{\sigma_s\}}2^{s-1}\sum_{l_{\sigma_1},...,l_{\sigma_s}=1}^{\infty}(-\xi)^{s+\sum_{r=1}^s l_{\sigma_r}}\exp\left[-\sum_{j=1}^{s}\alpha_{\sigma j}l_{\sigma_j}k^{\sigma_j}\right]\label{10},
\end{align}
with $ d $ being the number of compactified dimensions, $ \{\sigma_s\} $ represents the combination set with $ s $ elements and $ k $ is the 4-moment.

In the next sections, the ingredients presented here are used to investigate thermal and size effects in the electromagnetic theory modified by the aether term.

\section{The theory and its energy-momentum tensor}

Here the theoretical model is introduced and the energy-momentum tensor associated with it is calculated. The theory that describes the electromagnetic field coupled to the aether field, which violates Lorentz symmetries, is written in the thermal representation as
\begin{align}
	\widehat{\cal L}&={\cal L}-\tilde{\cal L}\nonumber\\
	&=-\frac{1}{4}F_{\mu\nu}F^{\mu\nu}-\frac{\kappa}{2}u^\mu u^\nu \eta^{\sigma\rho}F_{\mu\sigma}F_{\nu\rho}+\frac{1}{4}\tilde{F}_{\mu\nu}\tilde{F}^{\mu\nu}+\frac{\kappa}{2}u^\mu u^\nu \eta^{\sigma\rho}\tilde{F}_{\mu\sigma}\tilde{F}_{\nu\rho},
\end{align}
where $ {\cal L} $ and $ \tilde{\cal L} $ are the Lagrangians belonging to the original and tilde Hilbert spaces, respectively. The term responsible for breaking Lorentz symmetries is given by $ \kappa u^\mu u^\nu \eta^{\sigma\rho}F_{\mu\sigma}F_{\nu\rho} $, where $ \kappa$ is a coupling constant, $ u^\mu $ is a vector constant that acts as a background field and can be time-like or space-like and $F_{\mu\nu}$ is the standard electromagnetic tensor.

In order to write the energy-momentum tensor for this theory, its definition associated with the non-tilde field is considered, i.e., 
\begin{eqnarray}
T^{\mu\nu}=\frac{\partial{\cal L}}{\partial(\partial_\mu A^\lambda)}\partial^\nu A^\lambda-\eta^{\mu\nu}{\cal L}.\label{9}
\end{eqnarray}
It is important to note that there is a similar definition for the tilde field. Using Eq. (\ref{9}) the energy-momentum tensor associated with the electromagnetic field coupled to the aether field is
\begin{align}
	T^{\sigma\rho}=-F^\sigma_{\;\;\nu}\partial^\rho A^\nu+\frac{1}{4}\eta^{\sigma\rho}F_{\mu\nu}F^{\mu\nu}-\kappa(u^\mu u^\sigma F_{\mu\nu}-u^\mu u_\nu F^\sigma_\mu)\partial^\rho A^\nu+\frac{\kappa}{2}\eta^{\sigma\rho}u^\mu u^\nu \eta^{\lambda\delta}F_{\mu\lambda}F_{\nu\delta}.
\end{align}

To symmetrize this tensor, the Belinfante method is used. However, only the Lorentz invariant part becomes symmetric. This is a natural characteristic of Lorentz violating theories. Therefore, the Lorentz violating energy-momentum tensor is neither symmetric nor covariantly conserved. This characteristic has been found for different theories with Lorentz violation \cite{Kostelecky4, Kost, Mota}.  Then the energy-momentum tensor reads
\begin{align}
		T^{\sigma\rho}=-F^\sigma_{\;\;\nu}F^{\rho\nu}+\frac{1}{4}\eta^{\sigma\rho}F_{\mu\nu}F^{\mu\nu}-\kappa(u^\mu u^\sigma F_{\mu\nu}-u^\mu u_\nu F^\sigma_\mu)\partial^\rho A^\nu+\frac{\kappa}{2}\eta^{\sigma\rho}u^\mu u^\nu \eta^{\lambda\delta}F_{\mu\lambda}F_{\nu\delta}.
\end{align}

To avoid divergences, the energy-momentum tensor is written at different space-time points. Then
\begin{align}
	T^{\sigma\rho}(x)=&\lim_{x'\to x}\tau{\bigg\{}-F^\sigma_{\;\;\nu}(x)F^{\rho\nu}(x')+\frac{1}{4}\eta^{\sigma\rho}F_{\mu\nu}(x)F^{\mu\nu}(x')+\kappa{\bigg[ }\frac{1}{2}\eta^{\sigma\rho}u^\mu u^\nu \eta^{\lambda\gamma}F_{\mu\lambda}(x)F_{\nu\gamma}(x')\nonumber\\&-u^\mu u^\sigma F_{\mu\nu}(x)\partial^\rho A^\nu(x')+u^\mu u_\nu F^\sigma_\mu(x)\partial^\rho A^\nu(x'){\bigg]}{\bigg\}}.\label{12}
\end{align}
Using the Coulomb gauge,
\begin{eqnarray}
	[A_\mu(x),\pi_\nu(x')]=i\left[\delta_{\mu\nu}-\frac{1}{\nabla^2}\partial_\mu\partial_\nu\right]\delta(x-x'),
\end{eqnarray}
where $ \pi_\nu(x)=\partial_0 A_\nu(x) $ is the conjugate momentum of $ A_\nu $, taking the vaccum expectation value of the energy-momentum tensor and the definition given in Eq. (\ref{Prop}), the physical energy-momentum tensor becomes
\begin{eqnarray}
{\cal T}^{\sigma\rho(ab)}(x;\alpha)&=&-i\lim_{x'\to x}{\bigg\{}\Gamma^{\sigma\rho}(x,x')-\kappa{\bigg[}\frac{1}{2}\eta^{\sigma\rho}u^\mu u^\nu \eta^{\gamma\lambda}\Sigma_{\mu\nu\gamma\lambda}(x,x')-u^\mu u^\sigma\Pi^{\;\;\;\rho}_\mu(x,x')\nonumber\\
&&+\eta_{\nu\varepsilon}u^\mu u^\varepsilon\Theta^{\sigma\nu,\quad\rho}_{\quad\mu}(x,x'){\bigg]}{\bigg\}}\overline{G}_0^{(ab)}(x-x';\alpha),\label{14}
\end{eqnarray}
with
\begin{eqnarray}
\Gamma^{\sigma\rho}(x,x')&=&2\left(\partial^\sigma\partial'^\rho-\frac{1}{4}\eta^{\sigma\rho}\partial^\nu\partial'_\nu\right),\\
	\Sigma_{\mu\nu\lambda\gamma}(x,x')&=&\eta_{\rho\lambda}\partial_\mu\partial'_\nu-\eta_{\lambda\nu}\partial_\mu\partial'_\gamma-\eta_{\mu\gamma}\partial_\lambda\partial'_nu+\eta_{\mu\nu}\partial_\lambda\partial'_\gamma,\\
		\Pi_\mu^{\;\;\;\rho}(x,x')&=&2\partial_\mu\partial'^\rho,\\
	\Theta^{\sigma\nu,\quad\rho}_{\quad\mu}(x,x')&=&\eta^{\sigma\nu}\partial_\mu\partial'^\rho-\eta^\nu_\mu\partial^\sigma\partial'^\rho,
\end{eqnarray}
and
\begin{eqnarray}
\overline{G}_0^{(ab)}(x-x';\alpha)=G_0^{(ab)}(x-x;\alpha)-G_0^{(ab)}(x-x').
\end{eqnarray}
It should be noted that to write a finite energy-momentum tensor as given in Eq. (\ref{14}) a renormalization procedure has been performed, i.e.
\begin{eqnarray}
	{\cal T}^{\sigma\rho(ab)}(x;\alpha)=\langle T^{\sigma\rho(ab)}(x;\alpha) \rangle- \langle T^{\sigma\rho(ab)}(x) \rangle,
\end{eqnarray}
where $\langle T^{\sigma\rho(ab)}(x) \rangle$ is the vacuum expectation value of the energy-momentum tensor.

Now let's investigate what physical consequences Eq. (\ref{14}) leads to when considering different choices of the parameter $\alpha$ and different directions for the 4-vector $ u^\mu $.

\section{Thermal and size effect applications}

In this section, the topological structure of the TFD formalism is used. Then different phenomena such as thermal and size effects are treated on an equal footing.  

\subsection{Stefan-Boltzmann law for the electromagnetic field coupled to the aether field}

To calculate the Stefan-Boltzmann law associated with this theory, the topology $\Gamma_4^1=\mathbb{S}^1\times\mathbb{R}^{3}$ is considered. In this case $ \alpha=(\beta,0,0,0) $  is taken which implies that  the temporal dimension is compactified in a circle of length  $\beta$. For this choice the Bogoliubov transformation Eq. (\ref{10}) is given as
\begin{align}
	v^2(\beta)=\sum_{l_0=1}^\infty e^{-\beta k^0l_0},
\end{align}
and the Green function is
\begin{align}
	\overline{G}_0(x-x';\beta)=2\sum_{l_0=1}^\infty G_0(x-x'-i\beta l_0n_0),
\end{align}
with $ n_0^\mu=(1,0,0,0) $. With these ingredients, let us analyze the energy-momentum tensor for different directions of $ u^\mu $.

\subsubsection{Time-like constant vector}

Here, the constant vector has the form $ u^\mu=(1,0,0,0) $. To calculate the temperature-dependent energy density, let us consider $ \sigma=\rho=0 $. For this case, the energy-momentum tensor Eq. (\ref{14}) becomes
\begin{align}
	{\cal T}^{00(11)}(\beta)=-2i\lim_{x'\to x}\sum_{l_0=1}^{\infty}\left\{\frac{(1+\kappa)}{2}\left(3\partial^0\partial'^0+\partial^1\partial'^1+\partial^2\partial'^2+\partial^3\partial'^3\right)\right\}G_0(x-x'-i\beta l_0n_0),
\end{align}
that leads to
\begin{align}
	{\cal T}^{00(11)}(\beta)=\frac{\pi^2(1+\kappa)}{15\beta^4}\label{36},
\end{align}
which is exactly the Stefan-Boltzmann law for the electromagnetic field containing the contribution of the aether term that leads to the violation of Lorentz symmetries. For $ \sigma=\rho=3 $ the energy-momentum tensor is given as
\begin{align}
	{\cal T}^{33(11)}(\beta)&=-2i\lim_{x'\to x}\sum_{l_0=1}^{\infty}{\bigg\{}\frac{1}{2}\left(\partial^0\partial'^0-\partial^1\partial'^1-\partial^2\partial'^2+3\partial^3\partial'^3\right)\nonumber\\&+\frac{\kappa}{2}\left(3\partial^0\partial'^0-\partial^1\partial'^1-\partial^2\partial'^2+\partial^3\partial'^3\right){\bigg\}}G_0(x-x'-i\beta l_0n_0),
\end{align}
from which the pressure between photons is obtained as
\begin{align}
	{\cal T}^{33(11)}(\beta)=\frac{\pi^2(1+2k)}{45\beta^4}\label{38}.
\end{align}
It is observed that the aether term that violates the Lorentz symmetries contributes to increasing the pressure between the photons.

\subsubsection{Space-like constant vector} 

Considering the constant vector $u^\mu$ as space-like, three different cases are analyzed: (i) $ u^\mu=(0,1,0,0) $, (ii) $ u^\mu=(0,0,1,0) $ and (iii) $ u^\mu=(0,0,0,1) $. First, let us take $ u^\mu=(0,1,0,0) $. In this case, the energy-momentum tensor becomes
\begin{align}
	{\cal T}^{00(11)}(\beta)&=-2i\lim_{x'\to x}\sum_{l_0=1}^{\infty}{\bigg\{}\frac{1}{2}\left(3\partial^0\partial'^0+\partial^1\partial'^1+\partial^2\partial'^2+\partial^3\partial'^3\right)\nonumber\\&-\frac{\kappa}{2}\left(\partial^0\partial'^0+3\partial^1\partial'^1+\partial^2\partial'^2+\partial^3\partial'^3\right){\bigg\}}G_0(x-x'-i\beta l_0n_0),
\end{align}
from which we get
\begin{align}
	{\cal T}^{00(11)}(\beta)=\frac{\pi^2\left(3-2\kappa\right)}{45\beta^4}.
\end{align}
This result shows that the factor that breaks the Lorentz symmetries contributes to diminishing the Stefan-Boltzmann law. It is important to note that the other directions of the vector $u^\mu$, i.e. $ u^\mu=(0,0,1,0) $ and $ u^\mu=(0,0,0,1) $, lead to the same result.

Now let us analyze the pressure for the choice  $ u^\mu=(0,1,0,0) $. The energy-momentum tensor reads
\begin{align}
	{\cal T}^{33(11)}(\beta)&=-2i\lim_{x'\to x}\sum_{l_0=1}^{\infty}{\bigg\{}\frac{1}{2}\left(\partial^0\partial'^0-\partial^1\partial'^1-\partial^2\partial'^2+3\partial^3\partial'^3\right)\nonumber\\&-\frac{\kappa}{2}\left(\partial^0\partial'^0-3\partial^1\partial'^1-\partial^2\partial'^2+\partial^3\partial'^3\right){\bigg\}}G_0(x-x'-i\beta l_0n_0),
\end{align}
that provides us
\begin{align}
	{\cal T}^{33(11)}(\beta)=\frac{\pi^2}{45\beta^4}.
\end{align}
It is important to note that in this direction the term that violates Lorentz symmetries has no influence on the result. The same result is found for $ u^\mu=(0,0,1,0) $. However, for $ u^\mu=(0,0,0,1) $ a different analysis is obtained. Such a situation leads to  
\begin{align}
	{\cal T}^{33(11)}(\beta)&=-2i\lim_{x'\to x}\sum_{l_0=1}^{\infty}{\bigg\{}\frac{1}{2}\left(\partial^0\partial'^0-\partial^1\partial'^1-\partial^2\partial'^2+3\partial^3\partial'^3\right)\nonumber\\&-\frac{\kappa}{2}\left(\partial^0\partial'^0-\partial^1\partial'^1-\partial^2\partial'^2+3\partial^3\partial'^3\right){\bigg\}}G_0(x-x'-i\beta l_0n_0),
\end{align}
which results in
\begin{align}
	{\cal T}^{33(11)}(\beta)=\frac{\pi^2(1-k)}{45\beta^4}.
\end{align}
From where it can be seen that the aether term in the direction $ u^\mu=(0,0,0,1) $ contributes to decreasing the pressure. Therefore, it is concluded that the constant vector $ u^\mu $ influences the Stefan-Boltzmann law and the pressure between the photons in different ways.  It is important to note that the energy condition is not changed due to the aether term. The weak energy condition requires that, for every future-pointing time-like vector $t^a$,
\bea
T_{ab}t^a t^b\geq 0.
\eea
Our results shown that this condition is satisfied in all cases studied since $\kappa$ is supposed to be much smaller than one.

\subsection{Casimir effect for the electromagnetic field coupled to the aether field at zero temperature}

To describe the Casimir effect at zero temperature using the topological structure of the TFD formalism the compactification parameter $\alpha$ is chosen as  $ \alpha=(0,0,0,i2d) $. In this case, the coordinate $z$ is compactified into a circle of length $ L=2d $, where $d$ is the distance between the plates. In this topology the Bogoliubov transformation is given as 
\begin{align}
	v^2(d)=\sum_{l_3=1}^{\infty}e^{-i2dk^3l_3},
\end{align}
so that the Green function becomes
\begin{align}
	\overline{G}_0(x-x';d)=2\sum_{l_3=1}^{\infty}G_0(x-x'-2dl_3n_3),
\end{align}
where $ n_3=(0,0,0,1) $. With these elements, the Casimir effect at zero temperature for different directions of the constant vector $ u^\mu $ can be calculated.

\subsubsection{Time-like constant vector}

Considering the time-like vector $ u^\mu=(1,0,0,0) $, the component $ \sigma=\rho=0 $ of the energy-momentum tensor reads
\begin{align}
	{\cal T}^{00(11)}(d)=-2i\lim_{x'\to x}\sum_{l_3=1}^{\infty}{\bigg\{}\frac{(1+\kappa)}{2}\left(3\partial^0\partial'^0+\partial^1\partial'^1+\partial^2\partial'^2+\partial^3\partial'^3\right){\bigg\}}G_0(x-x'-2dl_3n_3),
\end{align}
from which we get
\begin{align}
	{\cal T}^{00(11)}(d)=-\frac{\pi^2(1+\kappa)}{720d^4}.
\end{align}
This is the Casimir energy at zero temperature.  It is verified that the factor that violates the Lorentz symmetries due to the aether term contributes to the Casimir energy associated with the electromagnetic field.

In order to calculate the Casimir pressure, the component $ \sigma=\rho=3 $ of the energy-momentum tensor is considered. Then
\begin{align}
	{\cal T}^{33(11)}(d)&=-2i\lim_{x\to x'}\sum_{l_3=1}^{\infty}{\bigg\{}\frac{1}{2}\left(\partial^0\partial'^0-\partial^1\partial'^1-\partial^2\partial'^2+\partial^3\partial'^3\right)\nonumber\\&+\frac{\kappa}{2}\left(3\partial^0\partial'^0-\partial^1\partial'^1-\partial^2\partial'^2+\partial^3\partial'^3\right){\bigg\}}G_0(x-x'-2dl_3n_3),
\end{align}
after some calculation,  the Casimir pressure is given as
\begin{align}
	{\cal T}^{33(11)}(d)=-\frac{\pi^2\left(3+2\kappa\right)}{720d^4}.
\end{align}
It can be seen that the Lorentz violation is an additive term, contributing to the attractive force between the plates.

\subsubsection{Space-like constant vector} 

Here let us analyze three different cases that emerge for  a space-like constant vector $ u^\mu $. Let's initially assume that it is $ u^\mu=(0,1,0,0) $ and thus the $ \sigma=\rho=0 $ component of the energy momentum tensor becomes
\begin{align}
	{\cal T}^{00(11)}(d)&=-2i\lim_{x'\to x}\sum_{l_3=1}^{\infty}{\bigg\{}\frac{1}{2}\left(3\partial^0\partial'^0+\partial^1\partial'^1+\partial^2\partial'^2+\partial^3\partial'^3\right)\nonumber\\&-\frac{\kappa}{2}\left(\partial^0\partial'^0+3\partial^1\partial'^1+\partial^2\partial'^2+\partial^3\partial'^3\right){\bigg\}}G_0(x-x'-2dl_3n_3),
\end{align}
that provides us
\begin{align}
	{\cal T}^{00(11)}(d)=-\frac{\pi^2}{720d^4}.
\end{align}
This is the Casimir energy associated with the electromagnetic field.  It is to be noted that the Casimir energy for this choice of vector $ u^\mu $ is not influenced by the violation of Lorentz symmetries. This is a natural consequence that arises from the choice of the vector $ u^\mu $.

In a similar way, the Casimir pressure for this choice is obtained as
\begin{eqnarray}
{\cal T}^{33(11)}(d)&=&-2i\lim_{x\to x'}\sum_{l_3=1}^{\infty}{\bigg\{}\frac{1}{2}\left(\partial^0\partial'^0-\partial^1\partial'^1-\partial^2\partial'^2+3\partial^3\partial'^3\right)\nonumber\\
&&-\frac{\kappa}{2}\left(\partial^0\partial'^0-3\partial^1\partial'^1-\partial^2\partial'^2+\partial^3\partial'^3\right){\bigg\}}G_0(x-x'-2dl_3n_3),\nonumber\\
&=&-\frac{\pi^2\left(3-2\kappa\right)}{720d^4}.
\end{eqnarray}
This result shows that the aether term changes the Casimir pressure, i.e., its contribution leads to a repulsive force between the plates, against the usual result which is attractive.  The same result is obtained for $ u^\mu=(0,1,0,0) $  and $ u^\mu=(0,0,1,0)$. However, for $ u^\mu=(0,0,0,1) $ the Casimir energy and pressure are given, respectively, as
\begin{eqnarray}
{\cal T}^{00(11)}(d)&=&-\frac{\pi^2(1-2\kappa)}{720d^4},\\
{\cal T}^{33(11)}(d)&=&-\frac{\pi^2(1-\kappa)}{240d^4}.
\end{eqnarray}
In both results, the aether term contributes to the reduction of the standard result for the electromagnetic field. It is important to emphasize that, assuming that the parameter that leads to the Lorentz violation is much smaller than one, then it does not change the behavior of the Casimir effect, i.e., it is always attractive.

\subsection{Casimir effect for the electromagnetic field coupled to the aether field at finite temperature}

In this section, the Casimir effect for the electromagnetic field at finite temperature is described, considering its interaction with the aether field. For such a description, it is necessary to compactify the temporal and spatial dimensions, so that $\alpha=(\beta,0,0,i2d)$. In this way, the Bogoliubov transformation is given as
\begin{align}
	v^2(\beta,d)=\sum_{l_0=1}^{\infty}e^{-\beta k^0l_0}+\sum_{l_3=1}^{\infty}e^{-i2dk^3l_3}+2\sum_{l_0,l_3=1}^{\infty}e^{-\beta k^0l_0-i2dk^3l_3},\label{BT}
\end{align}
where the first two terms describe the Stefan-Boltzmann law and the Casimir effect, respectively, while the last one describes both effects. The Green function corresponding to the last term of Eq. (\ref{BT}) is
\begin{align}
	\overline{G}_0(x-x';\beta,d)=4\sum_{l_0,l_3=1}^{\infty}G_0(x-x'-i\beta l_0n_0-2dl_3n_3).
\end{align}
As in the previous sections, using these quantities, the Casimir energy and pressure will be calculated for different directions of the constant vector $ u^\mu $.

\subsubsection{Time-like constant vector}

For $ u^\mu=(1,0,0,0) $, and considering the complete expression given in Eq. (\ref{BT}), the Casimir energy and pressure, after some calculations similar to those developed in previous sections, are given, respectively, as
\begin{eqnarray}
{\cal T}^{00(11)}(\beta,d)&=&\frac{\pi^2(1+\kappa)}{15\beta^4}-\frac{\pi^2(1+\kappa)}{720d^4}-\frac{4}{\pi^2}\sum_{l_0,l_3=1}^{\infty}(1+\kappa)\frac{\left[(2dl_3)^2-3(\beta l_0)^2\right]}{\left[(2dl_3)^2+(\beta l_0)^2\right]^3},\label{46}\\
{\cal T}^{33(11)}(\beta,d)&=&\frac{\pi^2(1+2k)}{45\beta^4}-\frac{\pi^2\left(3+2\kappa\right)}{720d^4}-\frac{4}{\pi^2}\sum_{l_0,l_3=1}^{\infty}\frac{(2dl_3)^2(2\kappa+3)-(\beta l_0)^2(2\kappa+1)}{\left[(2dl_3)^2+(\beta l_0)^2\right]^3}.\label{47}
\end{eqnarray}
The last term in these equations represent the thermal and size effects together. Note that the aether term changes all contributions.

Here it is interesting to investigate some temperature limits. First, let us consider the low-temperature limit, i.e., $\beta\gg 1$. Then the Casimir energy and pressure becomes
\begin{eqnarray}
{\cal T}^{00(11)}(\beta,d)&=&-(1+\kappa)\left[\frac{\pi^2}{720d^4}-\frac{\zeta(3)}{\pi\beta^3d}\right],\label{48}\\
{\cal T}^{33(11)}(\beta,d)&=&-\frac{\pi^2\left(3+2\kappa\right)}{720d^4}+\frac{\kappa\zeta(3)}{2\pi\beta^3d},\label{49}
\end{eqnarray}
where $\zeta(3)$ is the zeta function. Therefore, thermal effects impose a correction to the Casimir effect, although the negative nature of energy and pressure is dominant at this limit. Furthermore, if the high-temperature limit is applied to Eqs (\ref{46}) and (\ref{47}) it is noted that the positive contribution of the Stefan-Boltzmann law is dominant.

\subsubsection{Space-like constant vector} 

Here, the Casimir effect at finite temperature for the space-like constant vector is investigated. The choices  $ u^\mu=(0,1,0,0) $ and $ u^\mu=(0,0,1,0) $ lead to the same energy and pressure that are given as
\begin{eqnarray}
{\cal T}^{00(11)}(\beta,d)&=&\frac{\pi^2\left(3-2\kappa\right)}{45\beta^4}-\frac{\pi^2}{720d^4}-\frac{4}{\pi^2}\sum_{l_0,l_3=1}^{\infty}\frac{(2dl_3)^2-(\beta l_0)^2(3-2\kappa)}{\left[(2dl_3)^2+(\beta l_0)^2\right]^3},\\
{\cal T}^{33(11)}(\beta,d)&=&\frac{\pi^2}{45\beta^4}-\frac{\pi^2\left(3-2\kappa\right)}{720d^4}-\frac{4}{\pi^2}\sum_{l_0,l_3=1}^{\infty}\frac{(2dl_3)^2(3-2\kappa)-(\beta l_0)^2}{\left[(2dl_3)^2+(\beta l_0)^2\right]^3}.
\end{eqnarray}

Finally, taking $ u^\mu=(0,0,0,1) $ the energy and pressure at finite temperature are, respectively,
\begin{eqnarray}
{\cal T}^{00(11)}(\beta,d)&=&\frac{\pi^2\left(3-2\kappa\right)}{45\beta^4}-\frac{\pi^2}{720d^4}-\frac{4}{\pi^2}\sum_{l_0,l_3=1}^{\infty}\frac{(2dl_3)^2(1-2\kappa)-(\beta l_0)^2(3-2\kappa)}{\left[(2dl_3)^2+(\beta l_0)^2\right]^3},\\
{\cal T}^{33(11)}(\beta,d)&=&\frac{\pi^2(1-k)}{45\beta^4}-\frac{\pi^2(1-\kappa)}{240d^4}-\frac{4}{\pi^2}\sum_{l_0,l_3=1}^{\infty}(1-\kappa)\frac{\left[3(2dl_3)^2-(\beta l_0)^2\right]}{\left[(2dl_3)^2+(\beta l_0)^2\right]^3}.
\end{eqnarray}
It is important to note that contributions due to the aether term change the Casimir energy and pressure at finite temperature in different ways, depending on the direction of the four-vector $u^\mu$. These results for the space-like constant vector exhibit similar behavior to those shown in Eqs. (\ref{48}) and (\ref{49}) for high and low temperatures limits.

\section{Conclusion}

TFD is a real-time quantum field theory that exhibits an interesting topological structure. Here this formalism is used to study some applications involving thermal and size effects on a background that breaks Lorentz symmetry. Considering the aether field coupled to the electromagnetic field, the Stefan-Boltzmann law and the Casimir effect at zero and finite temperature are calculated. Our results show that both phenomena are affected by the aether term. Furthermore, the contribution of the Lorentz violation depends on the direction along which the aether field points. It is analyzed that the aether four-vector can be space-like or time-like. Then different choices lead to different results for both the Stefan-Boltzmann law and the Casimir effect. Note that the Casimir effect due to the aether field can be attractive or repulsive depending on the direction of the constant vector $u^\mu$. In addition, the limits at low and high temperatures are investigated. A comparison with the usual result for the Lorentz invariant electromagnetic field at zero temperature is performed. It is important to observe that in reference \cite{Our} this formalism is used to study the aether field in extra dimensions, where the aether field points exclusively along the extra direction. Therefore,  the analysis developed here is completely different, since the study is carried out in four-dimension and the aether field is considered in all directions. Furthermore, it is interesting to note that the results found here can be combined with cosmological and astrophysical data and then impose constraints on the Lorentz violation parameter. In the reference \cite{Manoel} a similar theory is considered and the Stefan-Boltzmann law is obtained. Using data from the CMB map, the Lorentz violation parameter is estimated. Therefore, studies such as the one developed here are useful for estimating values for the Lorentz violation parameter that agree with the observations. In addition, high-precision experimental measurements for the Casimir effect are a great ally to study the Lorentz violation, in  search of the trace left by this violation. Then constraints on the parameters of the model can be obtained if the measurements accuracy improves significantly.

\section*{Acknowledgments}

This work by A. F. S. is partially supported by National Council for Scientific and Technological Develo\-pment - CNPq project No. 313400/2020-2. R.C. and L. H. A. R. F. thanks CAPES for the financial support.

\section*{Data Availability Statement}

No Data associated in the manuscript.


\global\long\def\link#1#2{\href{http://eudml.org/#1}{#2}}
 \global\long\def\doi#1#2{\href{http://dx.doi.org/#1}{#2}}
 \global\long\def\arXiv#1#2{\href{http://arxiv.org/abs/#1}{arXiv:#1 [#2]}}
 \global\long\def\arXivOld#1{\href{http://arxiv.org/abs/#1}{arXiv:#1}}


\end{document}